\documentclass[a4paper,11pt]{article}
\usepackage{pos}
\usepackage{aas_macros}

\title{Locating the $\gamma$-ray emitting region in the quasar 4C\,+01.28}

\author*[a]{F. R{\"o}sch}
\author[a]{M. Kadler}
\author[b]{E. Ros}
\author[c]{M. Gurwell}
\author[d,e]{T. Hovatta}
\author[f]{M. Kreter}
\author[b]{N. R. MacDonald}
\author[g]{A. C. S. Readhead}

\affiliation[a]{Institut f{\"u}r Theoretische Physik und Astrophysik, Universit{\"a}t W{\"u}rzburg, Emil-Fischer-Str. 31, 97074 W{\"u}rzburg, Germany}

\affiliation[b]{Max-Planck-Institut f{\"u}r Radioastronomie, Auf dem H{\"u}gel 69, 53121 Bonn, Germany}

\affiliation[c]{Center for Astrophysics - Harvard \& Smithsonian, 60 Garden Street, Cambridge, MA 02138, USA}

\affiliation[d]{Finnish Centre for Astronomy with ESO (FINCA), University of Turku, FI-20014 Turku, Finland}

\affiliation[e]{Aalto University Mets{\"a}hovi Radio Observatory, Mets{\"a}hovintie 114, FI-02540 Kylm{\"a}l{\"a}, Finland}

\affiliation[f]{Centre for Space Research, North-West University, Private Bag X6001, Potchefstroom 2520, South Africa}

\affiliation[g]{Owens Valley Radio Observatory, California Institute of Technology, Pasadena, CA 91125, USA}

\emailAdd{florian.roesch@physik.uni-wuerzburg.de}

\abstract{Determining the location of $\gamma$-ray emission in blazar jets is a challenging task. Pinpointing the exact location of $\gamma$-ray production within a relativistic jet can place strong constraints on our understanding of high-energy astrophysics and astroparticle physics. We present a study of the radio- and $\gamma$-bright flat-spectrum radio quasar (FSRQ) 4C\,+01.28 (PKS\,B1055+018) in which we try to pinpoint the emission site of several prominent GeV flares. This source shows prominent high-amplitude broadband variability on time scales ranging from days to years. We combine high-resolution VLBI observations and multi-band radio light curves over a period of around nine years. We can associate two bright and compact newly ejected jet components with bright flares observed by the \textit{Fermi}/LAT $\gamma$-ray telescope and at various radio frequencies. A cross-correlation analysis reveals the radio light curves systematically lag behind the $\gamma$-rays. In combination with the jet kinematics as measured by the VLBA, we use these cross-correlations to constrain a model in which the flares become observable at a given frequency when a plasma component passes through the region at which the bulk energy dissipation takes place at that frequency. We derive a lower limit of the location of the $\gamma$-ray emitting region in 4C\,+01.28 of several parsecs from the jet base, well beyond the expected extent of the broad-line region. This observational limit challenges blazar-emission models that rely on the broad-line region as a source of seed photons for inverse-Compton scattering.}

\FullConference{%
  *** European VLBI Network Mini-Symposium and Users' Meeting (EVN2021) ***\\
  *** 12-14 July, 2021 ***\\
  *** Online ***
}

\begin{document}
\maketitle

\section{Introduction}
\noindent
Blazars are active galactic nuclei (AGN) that  emit radiation throughout the entire electromagnetic spectrum from radio frequencies up to high $\gamma$-ray energies. This high-energy $\gamma$-ray emission shows high-amplitude variability on very short time scales of less than a day \cite[e.g.,][]{HESS, Jorstad13}. While, the radio emission is thought to be produced by synchrotron radiation from relativistic electrons, there are two different classes of emission models to explain the high-energy $\gamma$-ray emission. On the one hand, the $\gamma$-ray emission is attributed to inverse Compton (IC) up-scattering of lower-energy seed photons. The seed photon field might be associated with the same relativistic electrons that are responsible for the synchrotron radiation or it might stem from external regions of the AGN \cite[e.g.,][]{Maraschi,Sikora09,Sikora94,Ghisellini09}. Therefore, the radio and $\gamma$-ray emission are expected to be correlated \cite{Lister4}, which is the reason why there are extensive studies of correlated radio-gamma properties \cite[e.g.,][]{Fuhrmann,Jorstad01,Max-Moerbeck,Kramarenko}. On the other hand, the high-energy $\gamma$-ray emission in AGN jets can be explained by hadronic emission models, in which pions are produced via photon-proton interactions within the jet and then decay in an electromagnetic particle cascade producing high-energy $\gamma$-ray photons and neutrinos \cite[e.g.,][]{Mannheim}. 

Blazars represent the largest population of extra-galactic objects in the $\gamma$-ray band \cite{Abdo1}. However, the exact location of the $\gamma$-ray emitting region is still under active discussion. The intraday variability shown by the $\gamma$-ray emission of several bright blazars indicates a compact emission region near the central super-massive black hole (SMBH). Therefore, the $\gamma$-ray emission of FSRQs (which exhibit strong optical emission-line spectra) is generally explained by external Compton (EC) scattering of broad-line region (BLR) photons \cite{Costamante}. But, in this case, there should be a strong cut-off at very high energies (VHE) due to $\gamma-\gamma$ absorption. However, no evidence for such a cut-off was found by analyzing the $\gamma$-ray spectra of 106 FSRQs \cite{Costamante}.  Moreover, multi-wavelength observations combined with Very Long Baseline Interferometry (VLBI) observations of FSRQs show newly ejected jet components that could be associated with bright $\gamma$-ray flares, which suggests that the $\gamma$-ray emitting region might be located well beyond the BLR \cite[e.g.,][]{Jorstad01}. 

The FSRQ 4C\,+01.28 (PKS\,B1055+018) is an ideal target to pinpoint the $\gamma$-ray emitting region within its jet, since the GeV $\gamma$-ray light
 curve\footnote{\textit{Fermi}/LAT 4FGL $\gamma$-ray light curve: \url{https://fermi.gsfc.nasa.gov/ssc/data/access/lat/8yr_catalog/}} shows high variability with bright flares. 
 Furthermore,  observations of 4C\,+01.28 with the Very Long Baseline Array (VLBA) at $43\,\mathrm{GHz}$\footnote{Boston University (BU) Blazar Monitoring Program: \url{https://www.bu.edu/blazars/VLBA_GLAST/1055.html}} show  bright and localized jet features well-suited for correlation studies. 
In the following we use a $\Lambda$CDM cosmological model with $H_{\mathrm{0}}=70\,\mathrm{km\,s^{-1}Mpc^{-1}}$, $\Omega_{\mathrm{m}}=0.30$ and $\Omega_{\mathrm{\Lambda}}=0.70$.

\section{43\,GHz VLBA observations}
\begin{figure}
    \centering
    \includegraphics[width=0.49\columnwidth]{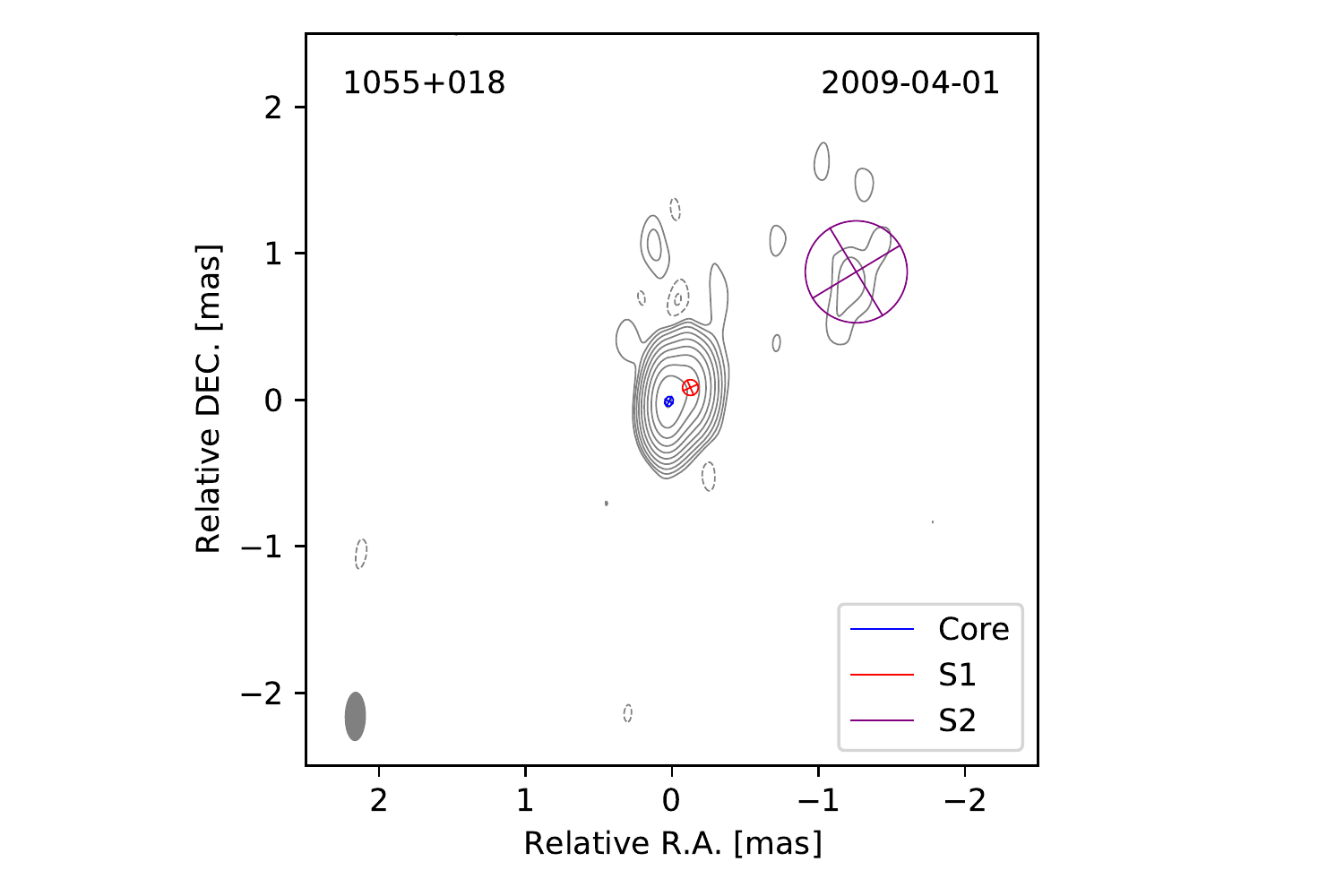}
    \includegraphics[width=0.49\columnwidth]{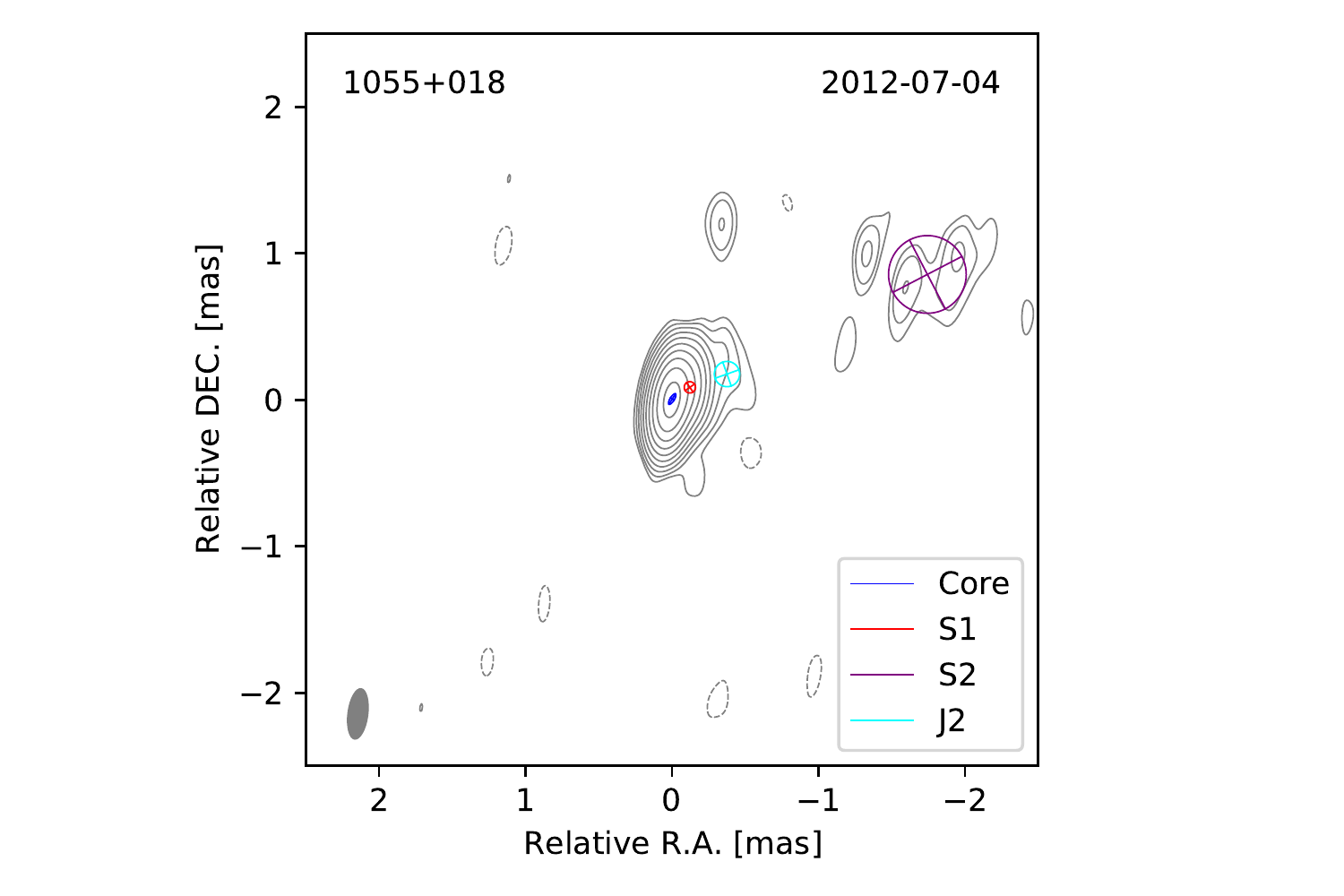}
    \includegraphics[width=0.49\columnwidth]{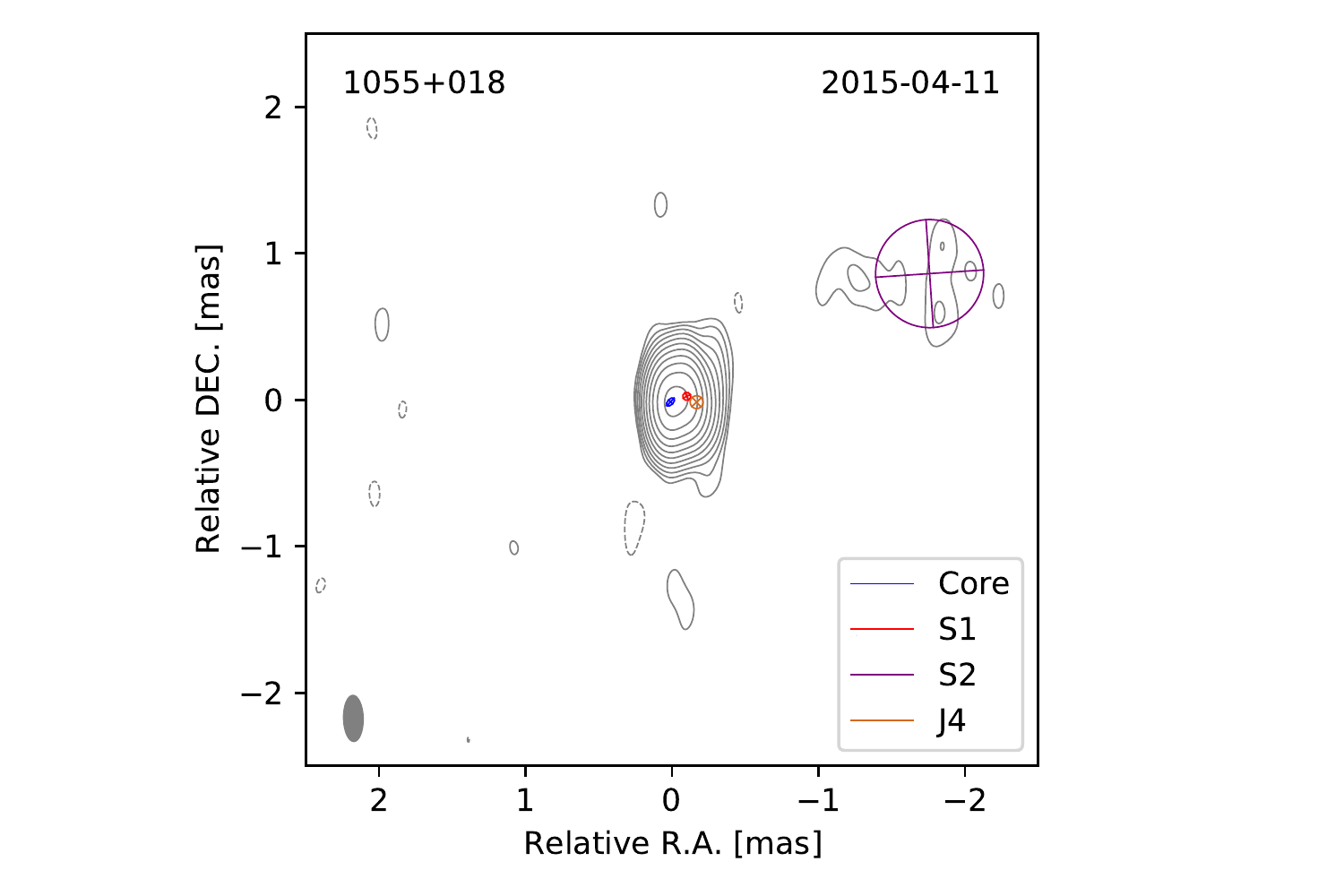}
    \includegraphics[width=0.49\columnwidth]{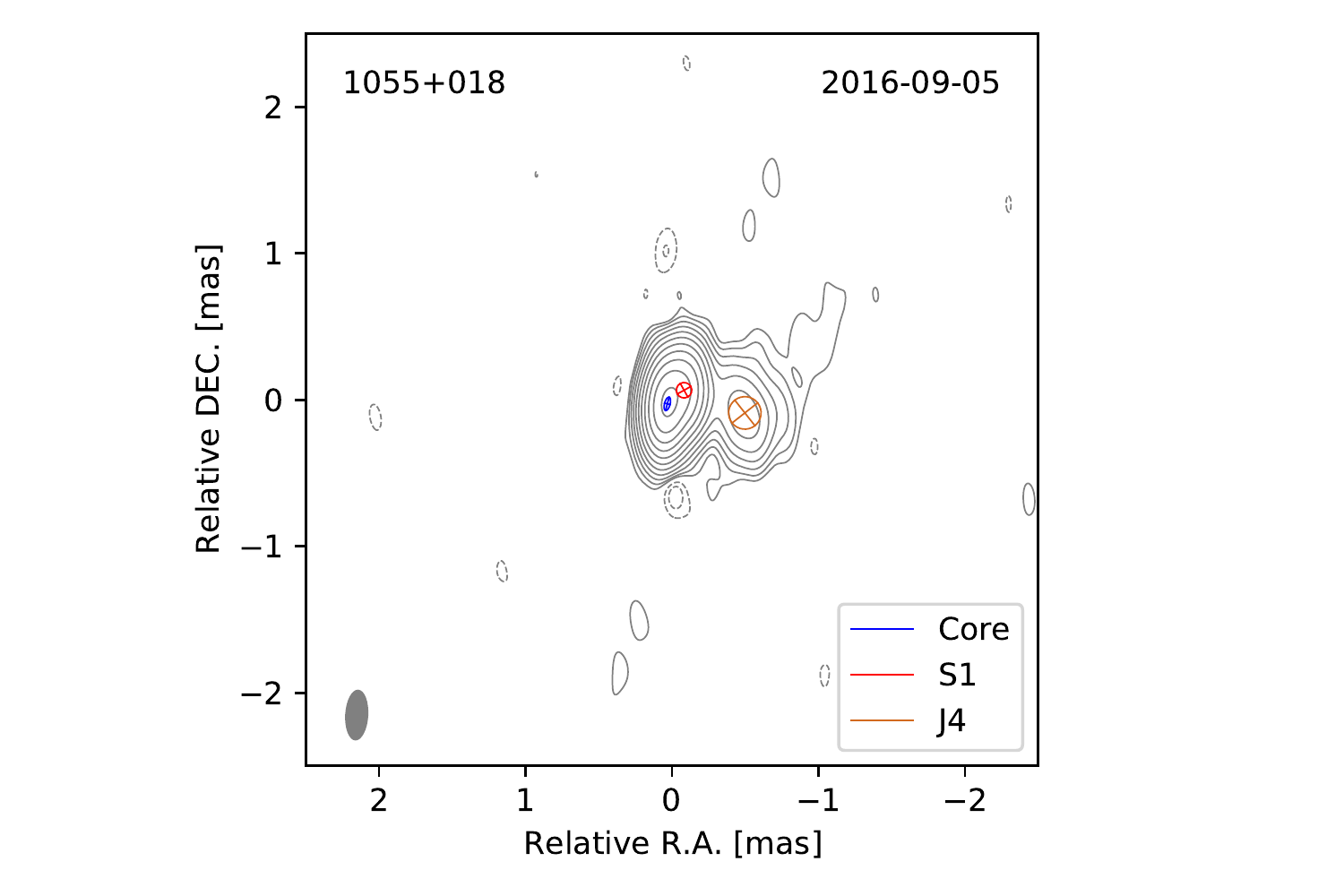}
    \includegraphics[width=0.49\columnwidth]{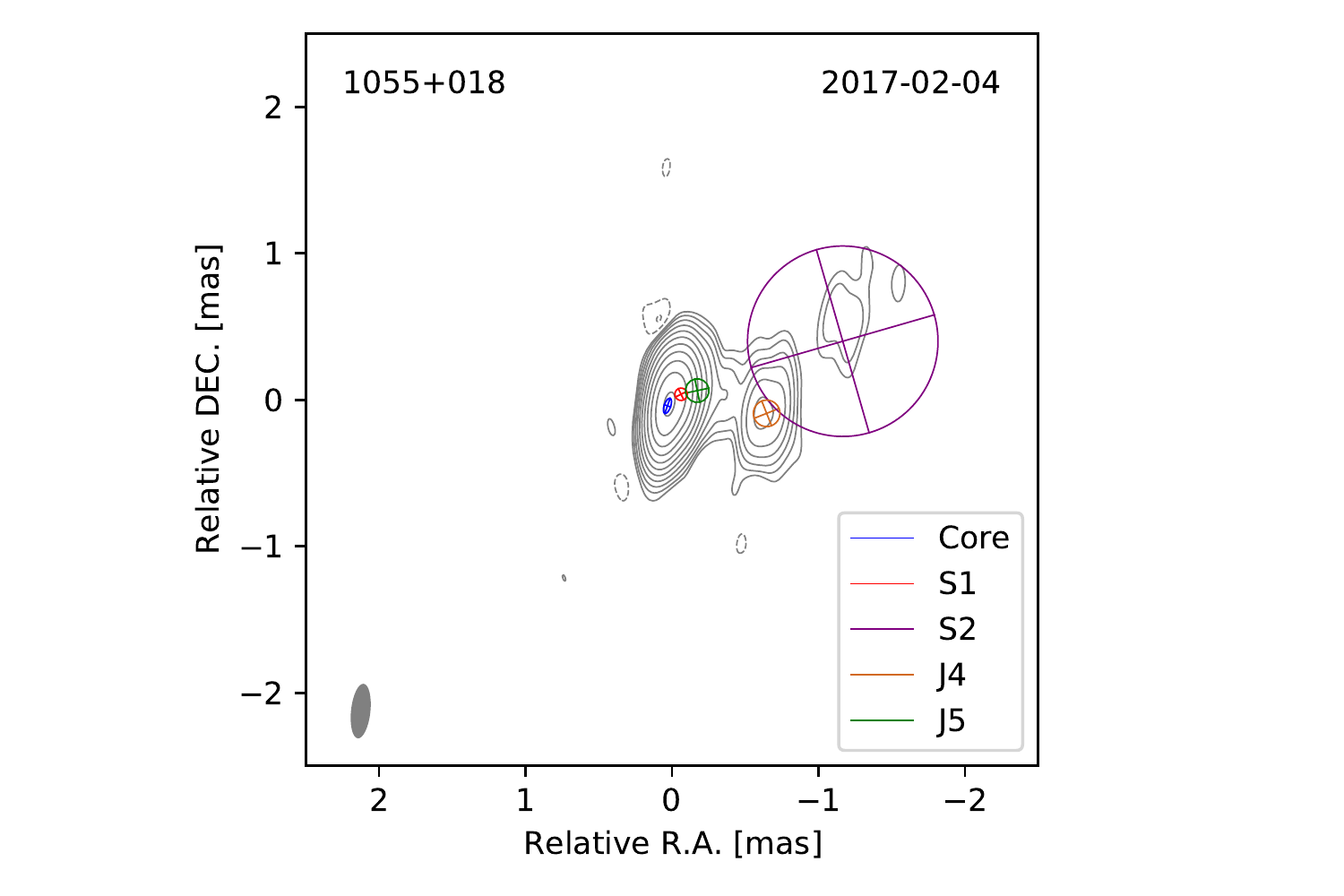}
    \includegraphics[width=0.49\columnwidth]{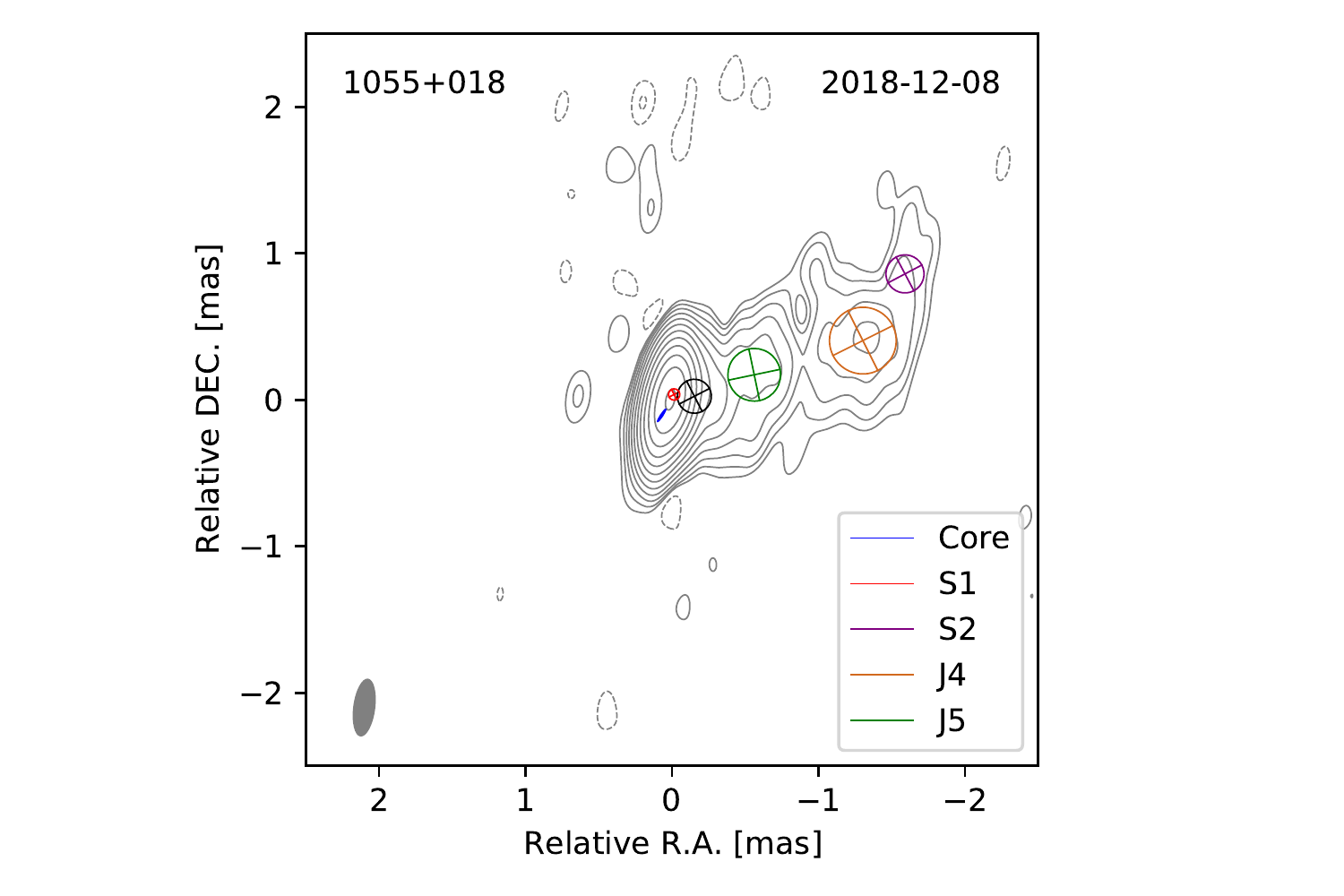}
    \caption{\small \sl Selected uniformly weighted images of $43\,\mathrm{GHz}$ VLBA observations of the FSRQ 4C\,+01.28 with  the fitted Gaussian components overlaid. The contours start at $3\sigma$ and increase logarithmically by factors of 2. The gray ellipse in the bottom left corner of each image corresponds to the beam. }
    \label{fig:komp}
\end{figure}

\noindent
To study the parsec-scale jet structure of the FSRQ 4C\,+01.28, we analyzed 51 epochs of VLBA observations at $43\,\mathrm{GHz}$ taken during a period of around nine years from April 2009 to December 2018. These data, provided by the BU Blazar Monitoring Program, had already been calibrated and imaged by the BU-group. More detailed information on the calibration and imaging process can be found in \cite{Jorstad05, Jorstad}. The uniformly weighted images of six selected epochs are shown in Fig.~\ref{fig:komp}.

To investigate the time evolution of the parsec-scale jet structure of 4C\,+01.28, we parametrized the fully calibrated and imaged data with 2D Gaussian components using the program \mbox{DIFMAP \cite{Shepherd2}}. We checked the resolution limit of these fitted components by calculating the lowest resolvable size as described in \cite{Lobanov}. Whenever a fitted axis of a component was found to be smaller than the corresponding resolution limit, we considered that value as an upper limit for the component's size. 

The resulting fitted Gaussian components overlaid on the uniformly weighted images are shown in Fig.~\ref{fig:komp} for six selected epochs of 4C\,+01.28. 
The bright central region can be fitted in most epochs with two or three components: the eastern most and brightest component of those is considered to be the core, while  one nearby additional compact component (S1) is considered to be an inner feature of the jet. In some epochs, additional jet components (J1, J2, J3; see discussion on their cross-epoch association below) are found in the same region or at distances out to about 0.6\,mas from the core.
One outer extended jet feature (S2) is persistently found in almost all epochs at distances between 1.5\,mas and 2\,mas west-northwest of the core.
In April 2015 a new compact jet component (J4) appeared that moves outwards in the jet and can be seen as a prominent bright knot in images at later epochs. Furthermore, another new jet component (J5) emerged in February 2017. This component also travels outward in the jet and appears as a second bright knot in later epochs. 


To determine the speeds and ejection times of the jet components we fitted their distance $d$ to the core component by linear regression using $d(t) = d_{\mathrm{mid}}+\mu(t-t_{\mathrm{mid}})$, in which $t_{\mathrm{mid}}=(t_{\mathrm{max}}+t_{\mathrm{min}})/2$ is the midpoint of the time interval in which the corresponding jet component was detected, $d_{\mathrm{mid}}$ is the distance of this jet component to the core component at the time $t_{\mathrm{mid}}$ and $\mu$ is the angular speed. For the uncertainties of the components' distances we used the semi major axes of the corresponding Gaussian components or the corresponding resolution limit for unresolved components.

\begin{figure}
    \centering
    \includegraphics[width=0.49\columnwidth]{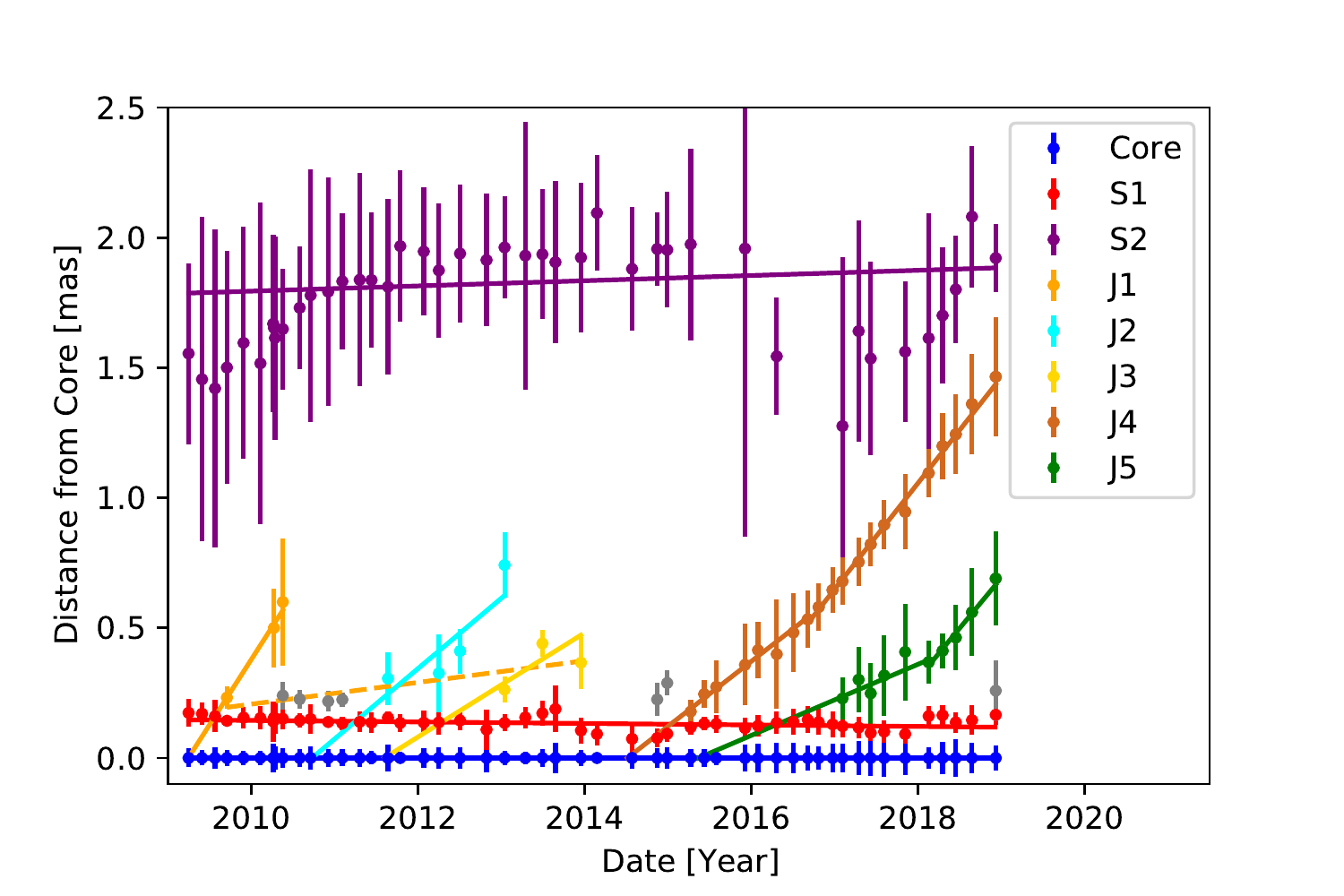}
    \includegraphics[width=0.49\columnwidth]{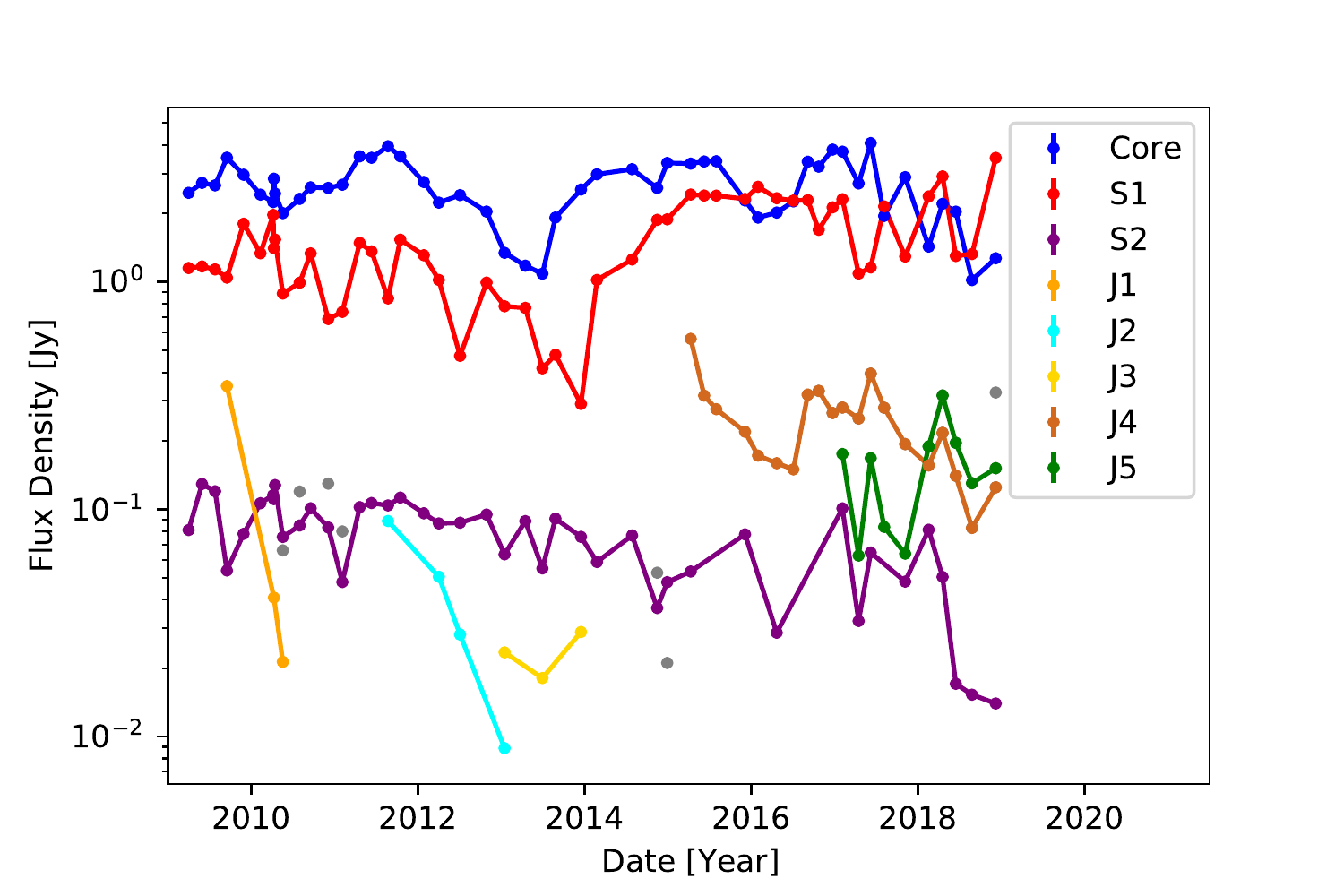}
    \caption{\small \sl Left panel: Distance of the jet components relative to the core component plotted over time. The solid lines were fitted by linear regression and their slopes represent the angular speed of the corresponding jet components. 
    The dashed line indicates a possible alternative cross-epoch association of model components (see text).
     Right panel: Flux density of the components as a function of time.}
    \label{fig:kin-flux}
\end{figure}

The distances of the jet components to the core component are plotted in Fig.~\ref{fig:kin-flux} (left panel). It is difficult to identify the jet components unambiguously at the beginning of the observation period. Therefore, we considered different models to describe the kinematics of these components. In a first model, we identified the third component within the bright feature at the center as one single component represented by the dashed line in the left panel of Fig.~\ref{fig:kin-flux}. This component then would move with an apparent speed of $\beta_{\mathrm{app}}=2.00\pm 0.61$, which would be much lower compared to the speeds computed for the newly ejected bright jet components J4 and J5 (see Table~\ref{tab:kin}). In a second model, we consider a scenario with three different jet components, J1, J2 and J3, represented by the solid lines in the left panel of Fig.~\ref{fig:kin-flux}. The speeds of these three components are more comparable to those of the components J4 and J5, but can be tracked only across three or four epochs. We therefore do not use their formal proper motions for further analysis\footnote{We adopt the convention of \cite[e.g.,][]{Lister1}, requiring at least five epochs to build a robust kinematics model.}. The two newly ejected jet components J4 and J5 seem to accelerate. Therefore, we fitted them by two separate linear regressions each, with their transition points located at distances of $\sim 0.53\,\mathrm{mas}$ for component J4 and $\sim 0.37\,\mathrm{mas}$ for component J5 with respect to the core component. We choose these transition points, because at that time when the components crossed their corresponding transition point their flux densities also show a steep increase, which can be seen in the right panel of Fig.~\ref{fig:kin-flux}, where the flux densities of the components are plotted with relative uncertainties of $5\%$. The speeds of the robust jet components are listed in Table~\ref{tab:kin}. 
Furthermore, we calculated the ejection epochs of the jet components J4 and J5 as that time where the separation of the corresponding jet component to the core component equals zero. These ejection epochs are also listed in Table~\ref{tab:kin}.

\begin{table*}
\centering
\small
\begin{tabular}{cccccc}
\hline\hline
Component & $d$ & $\mu$ & $\beta_{\mathrm{app}}$ & $t_{\mathrm{0}}$ & $\chi^2_{\mathrm{red}}$  \\
       & $[\mathrm{mas}]$ & $[\mathrm{mas\,yr^{-1}}]$ & $[c]$ & $[\mathrm{yr}]$  &         \\
       (1) & (2) & (3) & (4) & (5) & (6) \\
\hline
S1 & ----- & $-0.0029\pm0.0018$ & $-0.139\pm0.084$ & ----- & $0.30$	 	 \\
S2 & ----- & $0.010\pm0.014$ & $0.48\pm0.67$ & ----- & $0.34$	 	 \\
J4 & $\lesssim0.53$ & $0.250\pm0.066$ & $12.0\pm3.2$ & $2014.51\pm0.30$ & $0.045$	 	 \\
J4 & $\gtrsim0.53$ & $0.407\pm0.052$ & $19.5\pm2.5$ & ----- & $0.026$	 	 \\
J5 & $\lesssim0.37$ & $0.14\pm0.11$ & $6.5\pm5.1$ & $2015.4\pm1.7$ & $0.086$	 	 \\
J5 & $\gtrsim0.37$ & $0.39\pm0.21$ & $19\pm10$ & ----- & $0.026$	 	 \\
\hline
\end{tabular}
\caption{\label{tab:kin} \small \sl Apparent speeds and ejection epochs of the jet components. Col.(1): Component ID; Col.(2): Distance range (measured from the core) considered in the kinematics fit with respect to the kinematics transition point; Col.(3): Angular speed of the jet component; Col.(4): Apparent speed of the jet component in units of speed of light; Col.(5): Ejection epoch of the component; Col.(6): reduced $\chi^2$ value of the fit. Note that $\chi^2_{\mathrm{red}}<1$, due to the conservatively calculated uncertainties of the data points. In these cases, the uncertainties of the data points are larger than the mean deviation of these data points from the best fit.}
\end{table*}



We also analyzed the geometry of the jet of 4C\,+01.28. For this purpose, we assume that the diameter $D$ of the jet is well represented by the size of the Gaussian model components \cite[see, e.g.,][]{Burd2021}. Furthermore, we assume that $D\propto (d+d_\mathrm{c,\,43,\,app})^l$, where $d$ is the distance of the jet components to the core component, $d_\mathrm{c,\,43,\,app}$ is the apparent distance of the $43\,\mathrm{GHz}$ core component to the jet base and $l$ is a power law index defining the jet geometry \cite[see also][]{Kovalev20}. For a conical jet, $l=1$ \cite[e.g.,][]{Kadler}. We fitted the sizes of the jet components with respect to their distances to the core component via a power law fit of $D=C(d+d_{\mathrm{c,\,43,\,app}})^l$ leading to $d_\mathrm{c,\,43,\,app}=(0.112\pm0.056)\,\mathrm{mas}$ and $l=0.946\pm0.093$, which indicates a conical jet. Here, we considered only resolved jet components and assume a relative uncertainty of $20\%$ for the components' size.

\section{Multi-wavelength light-curve analysis}
\begin{figure}
    \vspace*{-0.8cm}
    \includegraphics[clip,width=0.7\textwidth]{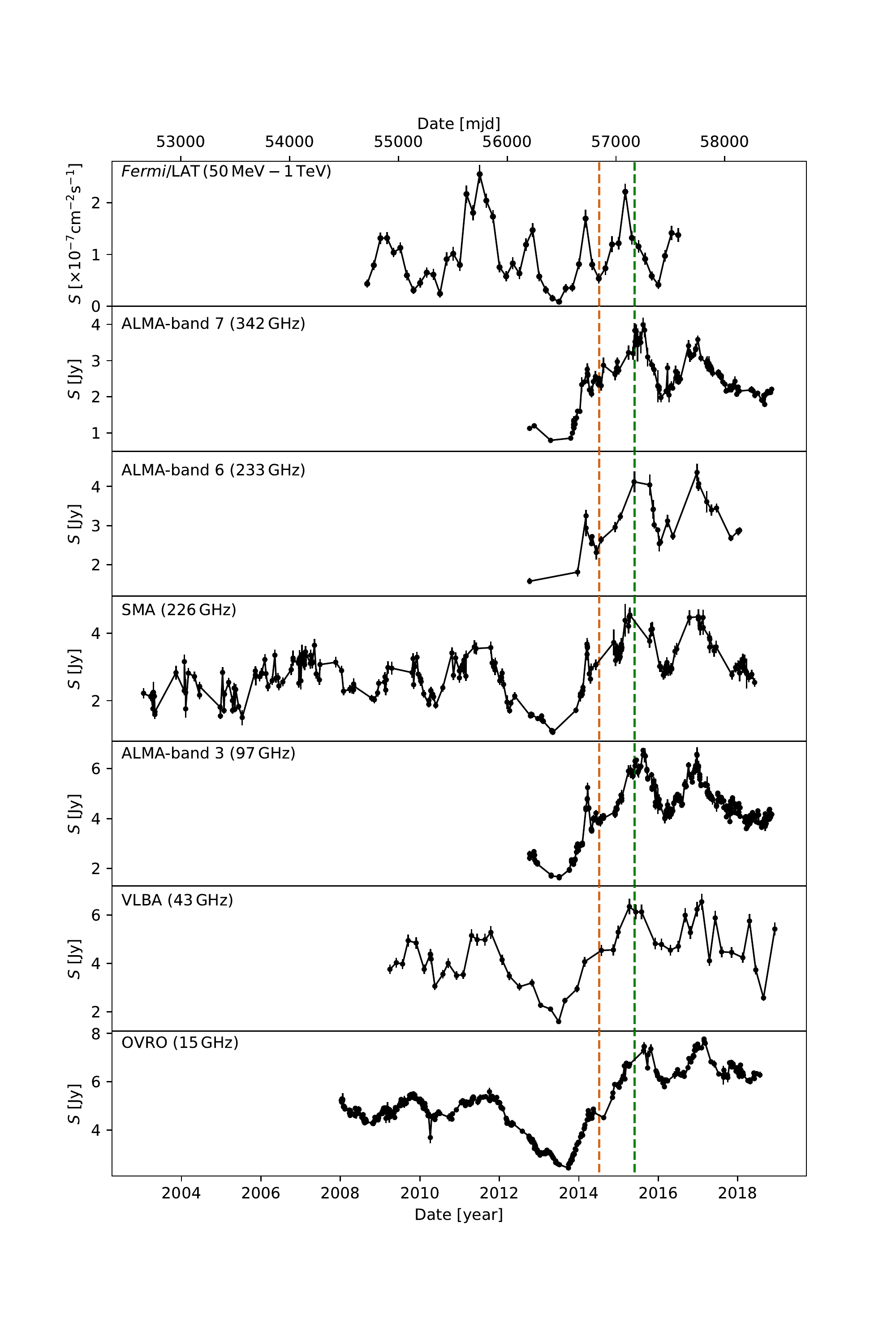}\hfill
    \vspace*{-7.5cm}
    
    \hfill\begin{minipage}{0.3\textwidth}
    \caption{\small \sl Two-month binned $\gamma$-ray and radio LCs observed by \textit{Fermi}/LAT, ALMA, SMA and OVRO and the total flux density VLBA LC, showing similar variability behavior. The given frequencies are the mean values of the observation frequencies of the individual LCs. The dashed lines represent the ejection times of J4 (brown) and J5 (green), respectively.}
    \label{fig:lc}
    \end{minipage}
\end{figure}

\noindent
Continuous information about the $\gamma$-ray brightness of
4C\,+01.28 can be obtained from observations done with the \textit{Fermi}/LAT telescope. The publicly available two-month binned 4FGL \textit{Fermi}/LAT $\gamma$-ray light curve (LC) is plotted in the upper panel of Fig.~\ref{fig:lc}. One can see that this LC shows high variability with two bright flares that can be associated with the ejection of the two bright jet components J4 and J5.

4C\,+01.28 was also observed by the Atacama Large Millimeter/submillimeter Array (ALMA) at band 3 (84 - 116 GHz), band 6 (211 - 275 GHz) and band 7 (275 - 373 GHz). These data are available at the ALMA Calibrator Source Catalogue\footnote{\url{https://almascience.eso.org/sc/}}. We used ALMA-data observed from October 6, 2012 to November 10, 2018. Furthermore, we also analyzed radio data measured by the Submillimeter Array (SMA) at a wavelength of $\lambda=1.3\,\mathrm{mm}$ from January 18, 2003 to June 5, 2018. 
These SMA-data are available at the SMA Submillimeter Calibrator List\footnote{\url{http://sma1.sma.hawaii.edu/callist/callist.html}}. Moreover, we also studied a LC measured by the Owens Valley Radio Observatory (OVRO) at 15 GHz from January 9, 2008 to July 23, 2018. All these different radio LCs are plotted in Fig.~\ref{fig:lc} together with the $\gamma$-ray LC and the total flux density 43 GHz VLBA LC. Since the ALMA and SMA LCs are observed at wide frequency ranges, we calculated the mean frequencies of the observations with uncertainties given by the standard deviation. These mean frequencies are given in Fig.~\ref{fig:lc} and Table~\ref{tab:corr} and were used for the further calculations. All radio LCs show similar behaviour to the $\gamma$-ray LC. Therefore, we performed a cross-correlation analysis between the \textit{Fermi}/LAT $\gamma$-ray LC and the various radio LCs. The ALMA-band 6 and VLBA LCs were neglected in this analysis due to their poor sampling rate compared to the other radio observations.

Because of the unevenly sampled radio LCs, we used two different methods for the cross-correlation analysis, namely the Discrete Cross-Correlation Function (DCF) \cite{Edelson} and the Interpolated Cross-Correlation Function (ICF) \cite{White}, combined with the bootstrap method introduced in \cite{Peterson} to compute uncertainties for the ICF time lags. We found positive time lags $\tau_{\mathrm{\gamma,\nu}}$ between the $\gamma$-ray LC and the different radio LCs which means that the $\gamma$-ray LC leads the radio LCs. The derived time lags and their corresponding maximum correlation coefficients are listed in Table~\ref{tab:corr}. 


\section{Locating the $\gamma$-ray emitting region}
\noindent
Using the results of the kinematic and cross-correlation analyses, we estimated a lower limit for the location of the $\gamma$-ray emitting region within the jet of 4C\,+01.28. To achieve this, we used a model presented in \cite{Max-Moerbeck} where the radio and $\gamma$-ray activity shown by the LCs plotted in Fig.~\ref{fig:lc} is produced when a moving jet component passes through the $\gamma$-ray emitting region at distance $d_{\mathrm{\gamma}}$ from the jet base and through the radio core regions at distances $d_{\mathrm{c,\,\nu}}$ from the jet base, respectively. Then, the apparent location of the $\gamma$-ray emitting region is given by $d_{\mathrm{\gamma,\,app}}=d_{\mathrm{c,\,43,\,app}}-\mu\cdot\tau_{\mathrm{\gamma,\,43}}$. However, we did not directly measure the time lag $\tau_{\mathrm{\gamma,\,43}}$ between the $\gamma$-ray and the 43 GHz VLBA LC. Therefore, we used the core shift to determine this time lag. The location of the radio core, where the optical depth becomes unity, shifts upstream in the jet with increasing frequency as $d_{\mathrm{c,\,\nu}}\propto\nu^{-\frac{1}{k_\mathrm{r}}}$ \cite{Konigl}. Assuming that the activity shown by the different radio LCs is produced by a moving jet component that traveled at a constant speed through the region of the jet where the radio cores are located, the core shift can be expressed by $\tau_{\mathrm{\gamma,\,\nu}}\propto\nu^{-\frac{1}{k_\mathrm{r}}}$ using the $\gamma$-ray emitting region as a reference position \cite{Fuhrmann}. Using this model, we determined the power law index $-\frac{1}{k_\mathrm{r}}$ by fitting the logarithms of the time lags as a function of the logarithms of the frequencies via linear regression which leads to $-\frac{1}{k_\mathrm{r}}=-0.12\pm0.34$ for the DCF time lags and $-\frac{1}{k_\mathrm{r}}=-0.25\pm0.95$ for the ICF time lags. With these power law indices the apparent location of the $\gamma$-ray emitting region can then be expressed by $d_{\mathrm{\gamma,\,app}}=d_{\mathrm{c,\,43,\,app}}-\mu\cdot\tau_{\mathrm{\gamma,\,\nu}}\cdot\left(43\,\mathrm{GHz}/\nu\right)^{-\frac{1}{k_\mathrm{r}}}$. 

\begin{table}
\centering
\small
\begin{tabular}{ccccc}
\hline\hline
Radio LC & $\nu$ & $r_{\mathrm{corr}}$ & $\tau_{\mathrm{\gamma,\nu}}$ & $d_{\mathrm{\gamma,\,app}}$ \\
     & $[\mathrm{GHz}]$ &  & $[\mathrm{d}]$ & $[\mathrm{mas}]$ \\
    (1) & (2) & (3) & (4) & (5) \\
\hline
\multicolumn{5}{l}{DCF:} \\  
    ALMA-band 7 & $341.9\pm8.5$ & $0.81\pm0.16$ & $120$ & $0.053\pm0.084$ \\
    SMA & $225.6\pm4.8$ & $0.71\pm0.12$ & $29$ & $0.098\pm0.058$ \\
    ALMA-band 3 & $97.0\pm6.2$ & $0.811\pm0.079$ & $128$ & $0.058\pm0.072$ \\
    OVRO & $15$ & $0.46\pm0.11$ & $96$ & $0.080\pm0.063$ \\
    \hline
    \multicolumn{5}{l}{ICF:} \\
    ALMA-band 7 & $341.9\pm8.5$ & $0.726$ & $105\pm95$ & $0.04\pm0.17$ \\
    SMA & $225.6\pm4.8$ & $0.594$ & $15\pm34$ & $0.103\pm0.061$ \\
    ALMA-band 3 & $97.0\pm6.2$ & $0.672$ & $128\pm102$ & $0.052\pm0.099$ \\
    OVRO & $15$ & $0.401$ & $8\pm314$ & $0.11\pm0.11$ \\
\hline
\end{tabular}
\caption{\label{tab:corr}\small \sl 
Results of the LC cross-correlation analysis
using the DCF (top) and ICF (bottom) methods. Col.(1): Radio LC; Col.(2): Mean frequency of the radio LC; Col.(3): Peak cross-correlation coefficient; Col.(4): Time lag corresponding to the peak cross-correlation coefficient; Col.(5): Apparent location of the $\gamma$-ray emitting region.}
\end{table}

With this formula we calculated the location of the $\gamma$-ray emitting region for all derived time lags separately, using the angular speed of component J5, since the cross-correlation seems to be driven by the flare associated with the ejection of this component. The derived values, listed in Table~\ref{tab:corr}, are consistent with each other within their uncertainties. Therefore, we calculated the weighted mean of the apparent location of the $\gamma$-ray emitting region to be $d_{\mathrm{\gamma,\,app}}=(0.081\pm0.027)\,\mathrm{mas}$. At a redshift of $z=0.89$ \cite{Jorstad} ($1\,\mathrm{mas}=7.77\,\mathrm{pc}$) this equals to $d_{\mathrm{\gamma,\,app}}=(0.63\pm0.21)\,\mathrm{pc}$. 


Finally, using the data presented in this work and considering an upper limit of the viewing angle of $\phi \leq 5^\circ$ derived from the VLBA data, we calculated a lower limit of the de-projected location of the $\gamma$-ray emitting region of $$d_{\mathrm{\gamma}} \gtrsim 4\,\mathrm{pc},$$ which is well beyond the expected extent of the BLR of $\lesssim1\,\mathrm{pc}$ \cite{Fuhrmann}. This limit is consistent with values derived for a small number of other sources
(\mbox{AO\,0235+164}: $d_{\mathrm{\gamma}}\geq(15\pm8)\,\mathrm{pc}$, and PKS\,1502+106:  $d_{\mathrm{\gamma}}=(12\pm9)\,\mathrm{pc}$; \cite{Max-Moerbeck}) and also consistent with sample studies \cite[e.g.,][]{Kramarenko}.

\section{Conclusion}
\noindent
We analyzed 51 epochs of 43 GHz VLBA observations of the quasar 4C\,+01.28 taken during a period of around nine years between April 2009 and December 2018. We found two newly ejected bright jet components that can be associated with two prominent flares shown by the two-month binned 4FGL \textit{Fermi}/LAT $\gamma$-ray LC and by several radio LCs measured at different frequencies. We performed a cross-correlation analysis between the $\gamma$-ray LC and the different radio LCs that results in positive time lags meaning that the radio LCs systematically lag behind the $\gamma$-rays. Combining the results of the kinematic and cross-correlation analyses, we calculated a lower limit of the location of the $\gamma$-ray emitting region with respect to the jet base of $d_{\mathrm{\gamma}} \gtrsim 4\,\mathrm{pc}$, using a model in which the flares shown by the LCs are produced when a moving jet component crosses the $\gamma$-ray emitting region and the different radio cores, respectively. We thus conclude that the $\gamma$-ray emitting region within the jet of 4C\,+01.28 is located far beyond the BLR. Therefore, EC scattering on BLR photons \cite[e.g.,][]{Sikora94}, which is commonly used to explain the $\gamma$-ray emission of FSRQs \cite{Costamante}, is disfavored in the case of 4C\,+01.28. Alternative models in which the $\gamma$-ray emission of 4C\,+01.28  is attributed to SSC scattering \cite[e.g.,][]{Maraschi} or to EC scattering on IR photons between the BLR and the dust torus \cite[e.g.,][]{Sikora09} as well as  EC scattering on photons from the cosmic microwave background (CMB) \cite[e.g.][]{Ghisellini09} remain vital options. Also hadronic emission models \cite[e.g.,][]{Mannheim} can be considered to explain the $\gamma$-ray emission in the jet of 4C\,+01.28.

\acknowledgments
\noindent
F.R. and M.K. acknowledge support from Deutsche Forschungsgemeinschaft grant DFG KA3252/4-1.
This study makes use of the following ALMA data: ADS/JAO.ALMA\#2011.0.00001.CAL. ALMA is a partnership of ESO (representing its member states), NSF (USA) and NINS (Japan), together with NRC (Canada), MOST and ASIAA (Taiwan), and KASI (Republic of Korea), in cooperation with the Republic of Chile. The Joint ALMA Observatory is operated by ESO, AUI/NRAO and NAOJ.
This study makes use of VLBA data from the VLBA-BU Blazar Monitoring Program (BEAM-ME and VLBA-BU-BLAZAR;
\url{http://www.bu.edu/blazars/BEAM-ME.html}), funded by NASA through the Fermi Guest Investigator Program. The VLBA is an instrument of the National Radio Astronomy Observatory, which is a facility of the National Science Foundation operated by Associated Universities, Inc.
The Submillimeter Array is a joint project between the Smithsonian Astrophysical Observatory and the Academia Sinica Institute of Astronomy and Astrophysics and is funded by the Smithsonian Institution and the Academia Sinica.

\normalsize
\bibliographystyle{JHEP}
\bibliography{bibtex.bib}


\end{document}